\documentclass{svproc}
\usepackage{url}

\usepackage{epsfig}
\usepackage{subfigure}
\usepackage{calc}
\usepackage{amssymb}
\usepackage{amstext} 
\usepackage{url}
\usepackage[table,xcdraw]{xcolor}
\usepackage[fleqn]{amsmath} 
\usepackage{pslatex}
\usepackage{enumitem}
\usepackage{textcomp}
\usepackage{array,multirow,graphicx}

\begin{document}

\title{
Towards the analysis of team members well-being 
}

\author{Zan Xu, 
Sari Nurfauziyyah, 
Anastasia Romanova, 
Kaamesh G S, 
Yiqun Gao, 
Maria Spichkova}
\authorrunning{Z. Xu et al.}

\institute{School of Computing Technologies, RMIT University, Melbourne, Australia\\
\email{maria.spichkova@rmit.edu.au}\\
}
\maketitle              

\abstract{%
Many recent research studies have focused on the well-being of software development team members, as this aspect may be critical not only for productivity and performance at work but also for the physical health and personal life of employees. Many studies agree that an important factor of team member well-being is whether team members feel appreciated and acknowledged for their contributions. 
This paper presents the results of a project on the team well-being analysis as well as the prototype developed within the project.
%
}


\section{\uppercase{Introduction}}
\label{sec:introduction}

Many recent research studies have focused on the well-being of software development team members.
The systematic literature review presented by Godliauskas and {\v{S}}mite~\cite{godliauskas2025well} highlighted that the topic became more prominent in the last decade. The trend is increasing. For example, between 2000 and 2003, the authors identified only three primary studies, while in the period from 2020 to 2023, they identified seventeen primary studies, indicating an increase of more than five times. The increase in interest in team members' well-being may be partially attributed to the issues observed in 2020/2021 related to COVID-19. The impact of working from home during the pandemic was discussed, for example, 
by Ford et al.~\cite{ford2021tale}, Ralph et al.~\cite{ralph2020pandemic}, Ju{\'a}rez-Ram{\'\i}rez et al.~\cite{juarez2022covid}, 
Uddin et al.~\cite{uddin2022qualitative}, 
Russo et al.~\cite{russo2021daily}, and 
Oliveira et al.~\cite{oliveira2020surveying}.

Since 2020, a larger percentage of software developers report feeling burned out. For example, comparing the annual reports conducted by Mind Share in 2019 and 2021, demonstrates that the percentage of workers who have experienced signs of poor mental wellbeing significantly increased, from 59\% to 76\%. 
Mental well-being might impact many aspects, ranging from physical health and personal life (see, e.g., the studies conducted by Tokdemir~\cite{tokdemir2022software}  and Grzywacz et al.~\cite{grzywacz2002work})
to productivity and performance at work (see, e.g., studies by
Graziotin et al.~\cite{graziotin2013happy} and Woo  et al.~\cite{woo2011impact}).

Many studies agree that an important factor of team member well-being is whether team members feel appreciated and acknowledged for their contributions, see, e.g., study by Montes et al.~\cite{montes2025factors}.  
However, there are currently only very few practical solutions to the issue. It might be very useful to have a tool-supported personalised framework to ensure the team members receive positive feedback and acknowledgement of their results. At the same time, the solution should not create a significant additional workload, as this itself may have a negative impact on well-being.

\emph{Contributions:}  
This paper presents the results of a project on the team's well-being analysis. 
We proposed a well-being platform, ZENith, which aim is to promote transparency around employee well-being by
encouraging weekly check-ins and peer recognition through so-called \emph{Kudos}. 
By capturing regular sentiment data, the platform could help identify early red flags, enabling managers and HR to address dissatisfaction proactively and maintain team well-being.

\section{\uppercase{Related Work}}
\label{sec:related}

\subsection{Perception of mental well-being in software development teams}

Wong et al.~\cite{wong2023mental} conducted interviews with software engineers to investigate their perceptions of mental well-being at work. The authors concluded that mental well-being at work should be analysed at three levels: individual, team, and organisational. 
Our solution goes across all three levels: it provides an organisation-based system supported by the organisational policies, impacts team dynamics as well as takes into account team members' personalities and individual needs.

Graziotin et al.~\cite{graziotin2013happy} investigated the potential correlation of affective states (emotions, moods, and feelings) of software developers and their self-assessed productivity.  
The authors concluded that the affective states of software developers are positively correlated with their self-assessed productivity. 
These results highlight the need to consider team well-being when planning the organisation's policies and corresponding activities. 

Greiler et al.~\cite{greiler2022actionable} 
proposed a framework for improving software developer experience. The authors conducted interviews and analysed developer experience factors, contextual characteristics (such as seniority, personal interests, company goals, etc), barriers to improvements, as well as strategies and coping mechanisms. Lack of incentives was identified as one of the barriers. The incentives might be of very different kinds, where one of them is the acknowledgement of their results and corresponding positive feedback.   

Montes et al.~\cite{montes2025factors} conducted a mixed-method cross-country study (a set of 15 interviews and a survey with 76 participants). The authors identified `support and recognition' as one of the core factors influencing the well-being of software developers. 

Allen and French~\cite{allen2023work}  conducted a literature review on work-family topics, and highlighted that the work-related mental well-being might affect the individual's private life and family.   

  Kurian and Thomas~\cite{kurian2023importance} investigated the impact of positive emotions on the performance of software developers. The authors conducted a literature review, the results of which highlighted that frequent experiences of positive emotions might have a positive impact on developers' performance.

Sampson~\cite{sampson2025enhancing} conducted semi-structured interviews to investigate what factors may influence software practitioners’
experiences and whether these factors influence the work-related performance. `Satisfaction and Well-being' was one of the core factors identified within this study. 

Overall, the related works highlight the mental well-being of software developers may have a significant impact on their productivity, while recognition and positive feedback on individual results might have a positive impact on the work-related mental well-being. 
%
  

\subsection{Teaching and learning aspects}

This project has been conducted within the Teaching and Leaning (T\&L) initiative 
proposed at the RMIT University, see \cite{spichkova2017autonomous,simic2016enhancing}. 
The aim of this initiative is to increase students' exposure to research-related activities and help them gain a better understanding of research methods within a team-based final year project course provided to Bachelor's and Master's students.  
There are a number of studies on this research area.
For example, a framework for teaching Computing
Research Methods was introduced by Holz at al.~\cite{holz2006research}.  
Some academics proposed to include research projects in the curriculum as an optional activity, e.g., as an additional summer activity, see~ 
Weinman et al.~\cite{weinman2015teaching}. 
Some academics proposed to include research experience as a compulsory course-based activity even on the undergraduate level, see, e.g., ~
Fernandez~\cite{fernandez2024generalizing}. Many studies suggested to include learning of research methiods in project-based courses, see, e.g., 
Host~\cite{host2002introducing}. 

To encourage the interest of students in research within Software Engineering on undergraduate and postgraduate levels of study, we suggest adding research components to the project-based final year courses as an optional bonus task. 
We propose having research components as a small activity, corresponding to approx. one week of a full-time study, i.e., to 35-40 hours of related activities. 
The topics of the research components align with those of the software development projects conducted within the semester as the core assessment activity. As the research components are optional, they aren't marked. 
Over the past years, this has led to a wide range of topics being covered by various teams, ranging from HCI and autonomous robotic systems (see results presented by Sun et al.~\cite{sun2018software} and Clunne-Kiely et al.~\cite{clunne2017modelling}) to the studies on visualisation or security aspects. 
For example, the research project conducted by George et al.~\cite{george2020usage} focused on visualising AWS usage. 
Zhao et al.~\cite{zhao2025visualisation} and Spichkova et al.~\cite{spichkova2020gosecure}  worked on the visualisation of the CIS benchmark scanning results and on the automation of scanning security vulnerabilities. 
The study conducted by Christianto et al.~\cite{christianto2017software} focused on visualising the data from vertical transport facilities.




\section{\uppercase{Proposed System}}
\label{sec:system}

We present the core elements of ZENith, a well-being monitoring system that aims to promote transparency around employee well-being. 
ZENith encourages weekly check-ins and peer recognition through so-called \emph{Kudos}, which can be provided at any time convenient to an employee. 
By capturing regular sentiment data, ZENith could help identify early red flags, enabling managers and HR to address well-being-related issues proactively. 

\begin{figure*}[ht!] 
\begin{center}
\includegraphics[width=\linewidth]{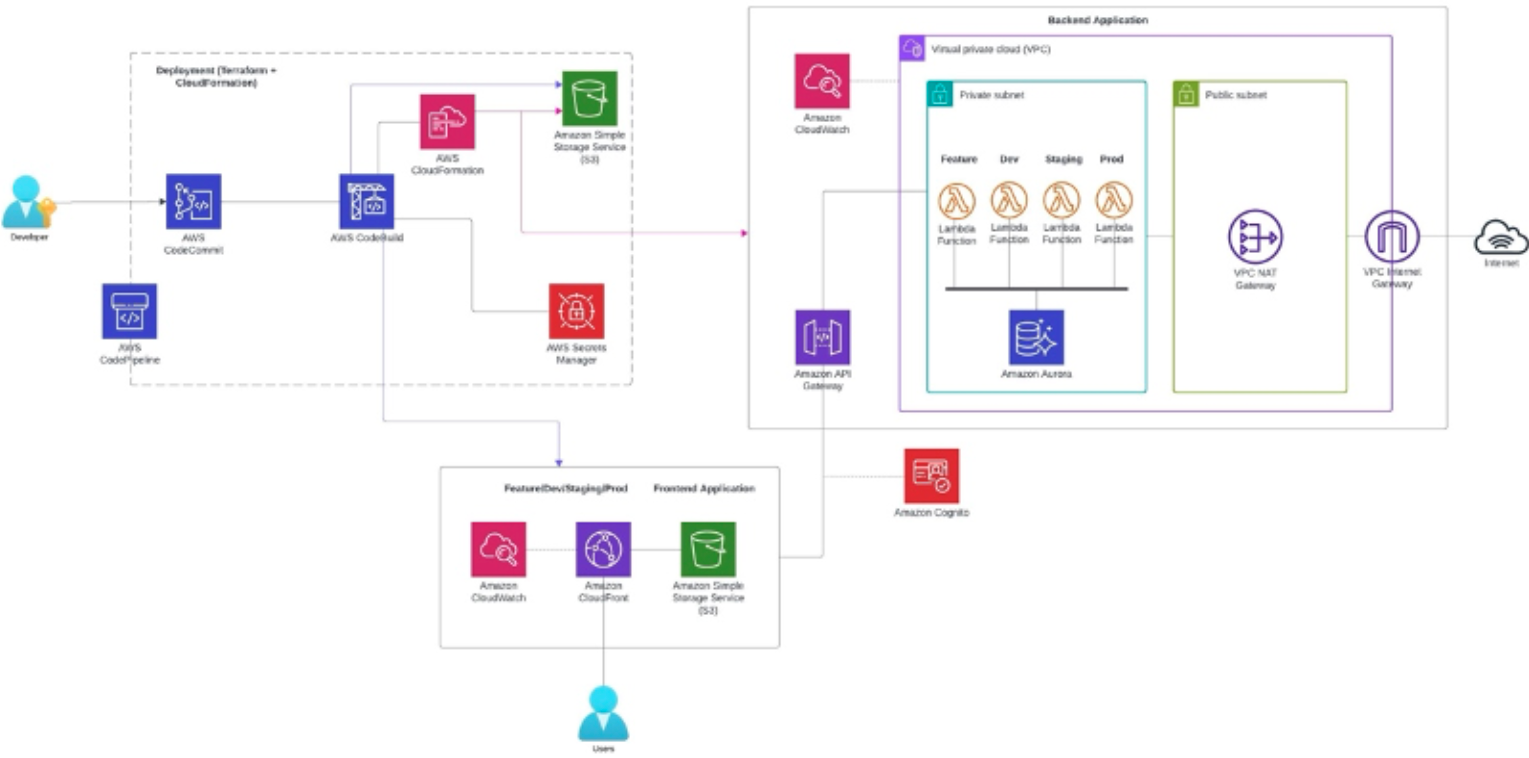}
\end{center}
\caption{Proposed system architecture}
\label{fig:arch2}
\end{figure*}

Figure~\ref{fig:arch2} presents the solution architecture for the proposed system. The architecture is event-driven and totally serverless, which guarantees scalability, affordability, and simplicity of implementation in a variety of settings.

The architecture includes:
\begin{itemize}
    \item An automated deployment pipeline via CodePipeline, CodeBuild, and CloudFormation/Terraform.
    \item  A frontend hosted on Amazon S3 and delivered via CloudFront.
    \item A backend consisting of API Gateway, Lambda functions, and Aurora
    \item MySQL, encapsulated within a VPC for security and scalability.
    \item Integration with Cognito for user authentication and CloudWatch for monitoring.
\end{itemize}

\begin{figure}[ht!] 
\begin{center}
\includegraphics[width=0.7\linewidth]{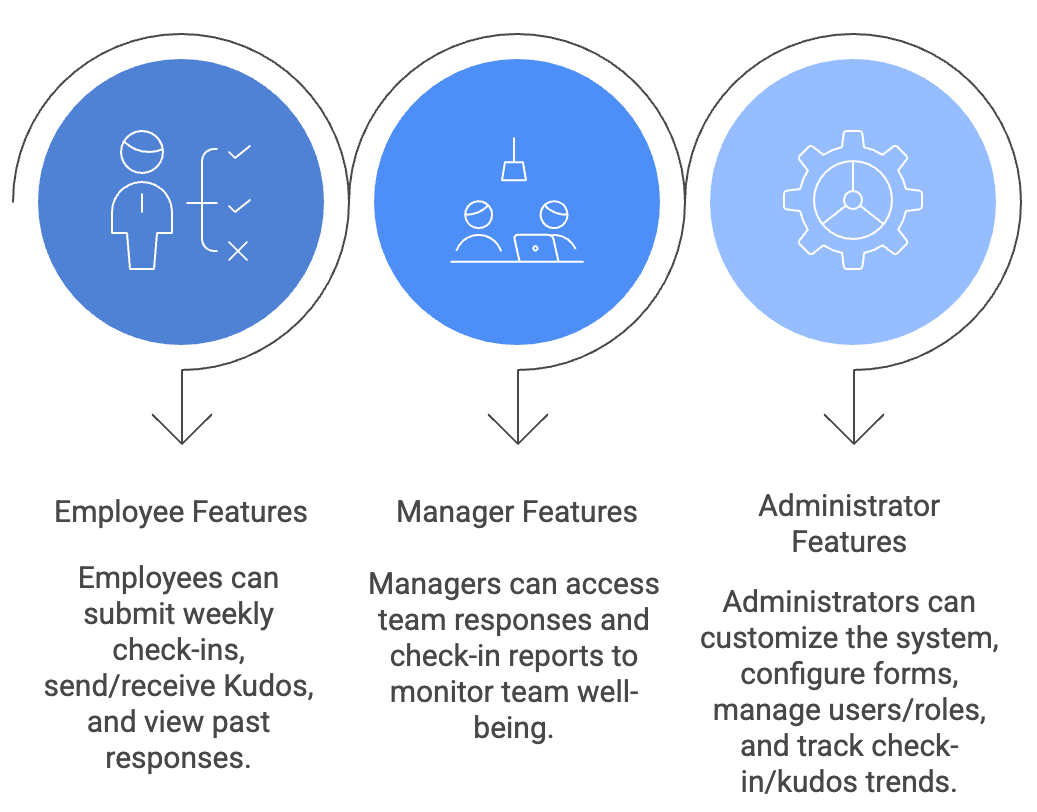}
\end{center}
\caption{ZENith: User types and features. (Note: To generate this figure from a manually created table, napkin.ai was
used.)}
\label{fig:users}
\end{figure}

\begin{figure}[ht!] 
\begin{center}
\includegraphics[width=0.8\linewidth]{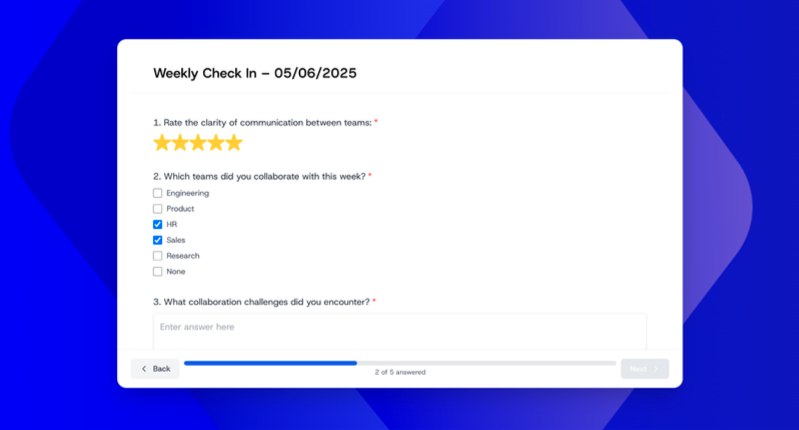}
\end{center}
\caption{Zenith: Filling out Weekly Check-in}
\label{fig:weekly}
\end{figure}

\begin{figure}[ht!] 
\begin{center}
\includegraphics[width=0.8\linewidth]{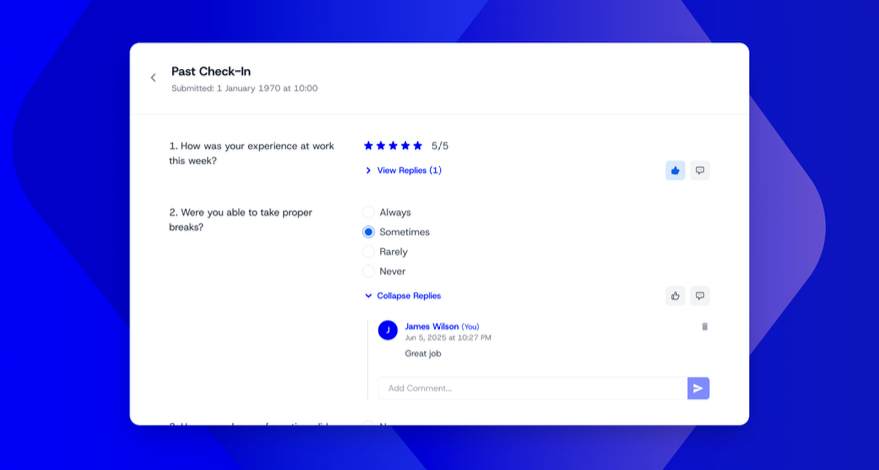}
\end{center}
\caption{Zenith: Commenting on specific check-ins}
\label{fig:commenting}
\end{figure}

\begin{figure}[ht!] 
\begin{center}
\includegraphics[width=0.6\linewidth]{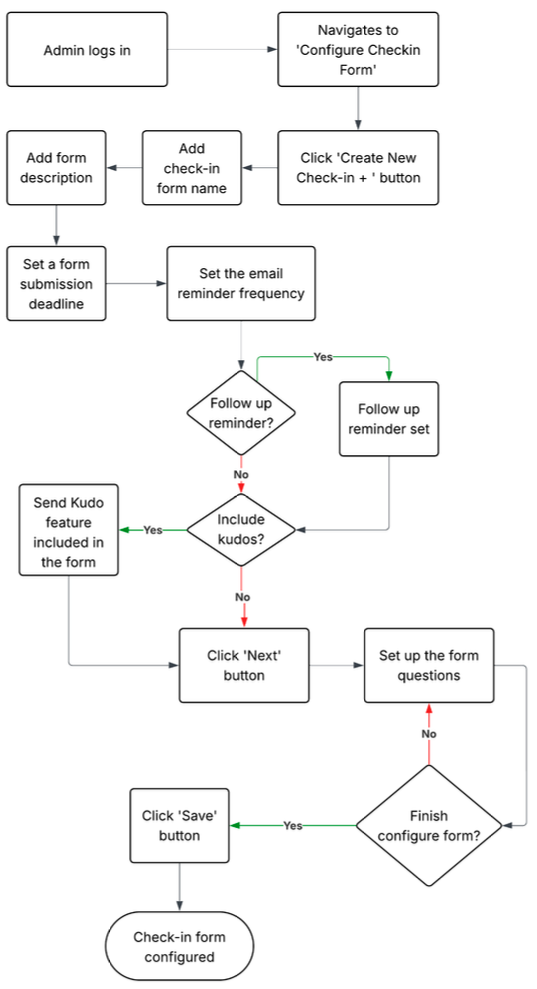}
\end{center}
\caption{Flow chart for the functionality `Check-In Configuration'}
\label{fig:configuration}
\end{figure}

ZENith provides 3 user types: Employees, Managers, and Administrators, see Figure~\ref{fig:users} for an illustrative overview of the corresponding features. The proposed system provides the following functionalities:
\begin{itemize}
    \item \emph{Employee Weekly Check-In} functionality enables employees to submit weekly reflections using a multi-question form, see Figure~\ref{fig:weekly} for a sample screenshot.
    \item \emph{Employee Kudos System} functionality allows employees to recognise each other for great work via a simple form.
    \item \emph{Manager \& Employee Comments} functionality allows users to comment on submitted answers, see Figure~\ref{fig:commenting} for a sample screenshot.
    \item \emph{Creation of a New Group.} Admins can create a new team group and assign members and managers.
    \item \emph{Editing of an Existing Group.} Admins can edit the existing team group and reassign members and managers.
    \item \emph{Check-In Configuration.} Admins can create and customise check-in forms and schedule email reminders, see Figure~\ref{fig:configuration} for the flow chart illustrating this functionality. 
\end{itemize}
 



\begin{figure}[ht!] 
\begin{center}
\includegraphics[width=0.8\linewidth]{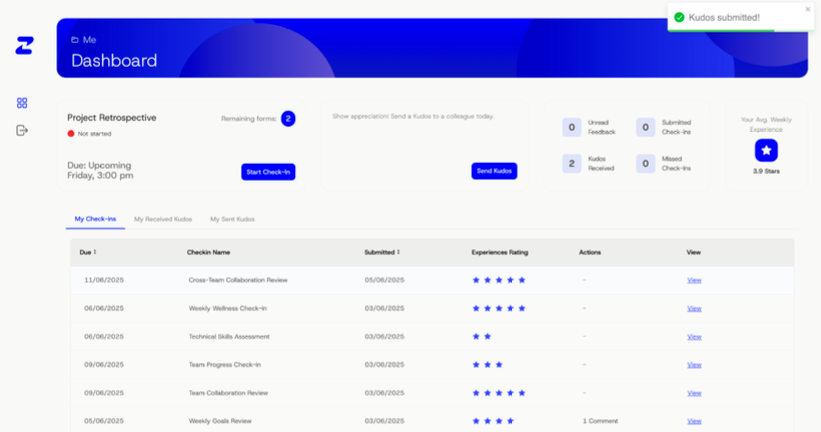}
\end{center}
\caption{Zenith: Manager dashboard}
\label{fig:dashboard}
\end{figure} 

The ZENith platform uses a relational database schema designed for clarity,
scalability, and seamless integration with Serverless infrastructure. The schema
supports user activity such as check-ins, kudos, feedback, and team management,
with relationships normalised across logical tables. Highlights include:
(1) Normalisation reduces redundancy and promotes maintainability.
(2) UUIDs are used as primary keys across all tables, ensuring globally unique identifiers and safe scalability.
(3) Role-based access enables secure differentiation of permissions (Employee, Manager, Admin).

The UI design is intuitive and accessible, with user-friendly multi-step check-in forms. We also took into account the findings of our prior research~\cite{alzahrani2022impact,alzahrani2021human,uitdenbogerd2022web} on the impact of interface elements on autistic and non-autistic users. 
Zenith was developed using React and Material UI components for a clean and consistent UX. 
We adopted  aria labels to provide textual descriptions for most UI elements, making Zenith more accessible to users who rely on assistive technologies like screen readers, see, e.g.,~\cite{cook2019assistive,borg2011right}.

\begin{figure}[ht!] 
\begin{center}
\includegraphics[width=0.8\linewidth]{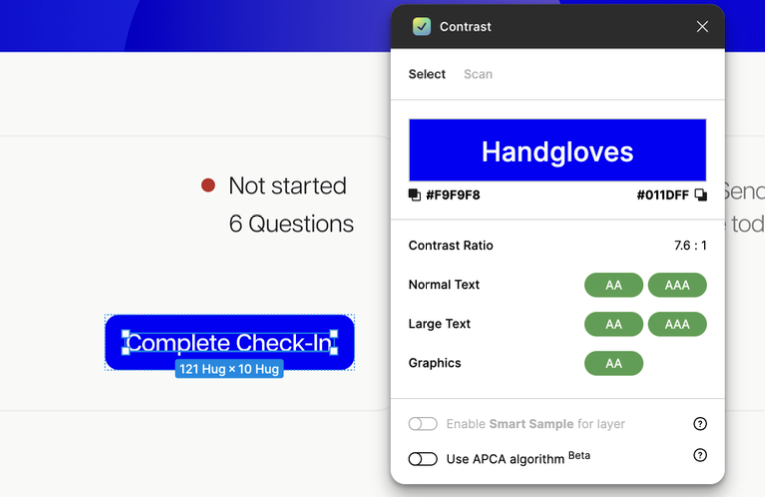}
\end{center}
\caption{Live CTA button with overlay confirming AAA compliance}
\label{fig:F3}
\end{figure}

The proposed system follows the Web Content Accessibility Guidelines (WCAG)~\cite{caldwell2008web}.  
Figure~\ref{fig:F3} illustrates how Zenith supports 7.6:1 colour contrast ratio, which exceeds the minimum requirements for web accessibility (WCAG AA) and meets the enhanced standard (WCAG AAA) for normal-sized text.

%

\section{\uppercase{Conclusions}} 
\label{sec:conclusions}

Recent research studies have increasingly focused on the well-being of software development team members, as this aspect is critical not only for workplace productivity and performance but also for employees' physical health and personal lives. 
Research consistently shows that team members' well-being is significantly influenced by their sense of appreciation and acknowledgment for their contributions.
In this paper, we presented the core results of a research project on the team well-being analysis.  The goal of the project was to develop a platform for analysing the team's well-being.  
We also present ZENith, a research prototype developed within the project in collaboration with Shine Solutions. 
%


\section*{{Acknowledgements}}
We would like to thank Shine Solutions for sponsoring this project under the research grant PRJ00002626, and especially Branko Minic and Adrian Zielonka for sharing their industry-based expertise and advice.

\bibliographystyle{splncs04}
 \bibliography{lit}

@article{ford2021tale,
  title={A tale of two cities: Software developers working from home during the covid-19 pandemic},
  author={Ford, Denae and Storey, Margaret-Anne and Zimmermann, Thomas and Bird, Christian and Jaffe, Sonia and Maddila, Chandra and Butler, Jenna L and Houck, Brian and Nagappan, Nachiappan},
  journal={ACM Transactions on Software Engineering and Methodology (TOSEM)},
  volume={31},
  number={2},
  pages={1--37},
  year={2021},
  publisher={ACM New York, NY}
}

@inproceedings{wong2023mental,
  title={Mental wellbeing at work: Perspectives of software engineers},
  author={Wong, Novia and Jackson, Victoria and Van Der Hoek, Andr{\'e} and Ahmed, Iftekhar and Schueller, Stephen M and Reddy, Madhu},
  booktitle={Proceedings of the 2023 CHI Conference on Human Factors in Computing Systems},
  pages={1--15},
  year={2023}
}

@article{allen2023work,
  title={Work-family research: A review and next steps},
  author={Allen, Tammy D and French, Kimberly A},
  journal={Personnel Psychology},
  volume={76},
  number={2},
  pages={437--471},
  year={2023},
  publisher={Wiley Online Library}
}

@article{montes2025factors,
  title={The Factors Influencing Well-Being in Software Engineers: A Cross-Country Mixed-Method Study},
  author={Montes, Cristina Martinez and Penzenstadler, Birgit and Feldt, Robert},
  journal={arXiv preprint arXiv:2504.01787},
  year={2025}
}

@article{woo2011impact,
  title={Impact of depression on work productivity and its improvement after outpatient treatment with antidepressants},
  author={Woo, Jong-Min and Kim, Won and Hwang, Tae-Yeon and Frick, Kevin D and Choi, Byong Hwi and Seo, Yong-Jin and Kang, Eun-Ho and Kim, Se Joo and Ham, Byong-Joo and Lee, Jun-Seok and others},
  journal={Value in Health},
  volume={14},
  number={4},
  pages={475--482},
  year={2011},
  publisher={Elsevier}
}

@article{greiler2022actionable,
  title={An actionable framework for understanding and improving developer experience},
  author={Greiler, Michaela and Storey, Margaret-Anne and Noda, Abi},
  journal={IEEE Transactions on Software Engineering},
  volume={49},
  number={4},
  pages={1411--1425},
  year={2022},
  publisher={IEEE}
}

@inproceedings{graziotin2013happy,
  title={Are happy developers more productive? The correlation of affective states of software developers and their self-assessed productivity},
  author={Graziotin, Daniel and Wang, Xiaofeng and Abrahamsson, Pekka},
  booktitle={International Conference on Product Focused Software Process Improvement},
  pages={50--64},
  year={2013},
  organization={Springer}
}

@inproceedings{sampson2025enhancing,
  title={Enhancing Developer Experience: Burnout, Resilience, and the Digital Environment in Software Engineering},
  author={Sampson, Michael G},
  booktitle={38th International BCS Human-Computer Interaction Conference},
  pages={369--374},
  year={2025},
  organization={BCS Learning \& Development}
}

@article{uitdenbogerd2022web,
  title={Web-based search: How do animated user interface elements affect autistic and non-autistic users?},
  author={Uitdenbogerd, Alexandra L and Spichkova, Maria and Alzahrani, Mona},
  journal={arXiv preprint arXiv:2211.11993},
  year={2022}
}

@article{kurian2023importance,
  title={Importance of positive emotions in software developers’ performance: a narrative review},
  author={Kurian, Riba Maria and Thomas, Shinto},
  journal={Theoretical Issues in Ergonomics Science},
  volume={24},
  number={6},
  pages={631--645},
  year={2023},
  publisher={Taylor \& Francis}
}

@article{grzywacz2002work,
  title={Work--family spillover and daily reports of work and family stress in the adult labor force},
  author={Grzywacz, Joseph G and Almeida, David M and McDonald, Daniel A},
  journal={Family relations},
  volume={51},
  number={1},
  pages={28--36},
  year={2002},
  publisher={Wiley Online Library}
}

@article{tokdemir2022software,
  title={Software professionals during the COVID-19 pandemic in Turkey: Factors affecting their mental well-being and work engagement in the home-based work setting},
  author={Tokdemir, Gul},
  journal={Journal of Systems and Software},
  volume={188},
  pages={111286},
  year={2022},
  publisher={Elsevier}
}

@inproceedings{oliveira2020surveying,
  title={Surveying the impacts of COVID-19 on the perceived productivity of Brazilian software developers},
  author={Oliveira Jr, Edson and Leal, Gislaine and Valente, Marco T{\'u}lio and Morandini, Marcelo and Prikladnicki, Rafael and Pompermaier, Leandro and Chanin, Rafael and Caldeira, Clara and Machado, Let{\'\i}cia and De Souza, Cleidson},
  booktitle={Proceedings of the XXXIV Brazilian Symposium on Software Engineering},
  pages={586--595},
  year={2020}
}

@inproceedings{russo2021daily,
  title={The daily life of software engineers during the covid-19 pandemic},
  author={Russo, Daniel and Hanel, Paul HP and Altnickel, Seraphina and Van Berkel, Niels},
  booktitle={2021 IEEE/ACM 43rd International Conference on Software Engineering: Software Engineering in Practice (ICSE-SEIP)},
  pages={364--373},
  year={2021},
  organization={IEEE}
}

@article{uddin2022qualitative,
  title={A qualitative study of developers’ discussions of their problems and joys during the early COVID-19 months},
  author={Uddin, Gias and Alam, Omar and Serebrenik, Alexander},
  journal={Empirical software engineering},
  volume={27},
  number={5},
  pages={117},
  year={2022},
  publisher={Springer}
}

@article{juarez2022covid,
  title={How COVID-19 pandemic affects software developers’ wellbeing, and the necessity to strengthen soft skills},
  author={Ju{\'a}rez-Ram{\'\i}rez, Reyes and Navarro, Christian X and Licea, Guillermo and Jim{\'e}nez, Samantha and Tapia-Ibarra, Ver{\'o}nica and Guerra-Garc{\'\i}a, C{\'e}sar and Perez-Gonzalez, Hector G},
  journal={Programming and Computer Software},
  volume={48},
  number={8},
  pages={614--631},
  year={2022},
  publisher={Springer}
}

@article{ralph2020pandemic,
  title={Pandemic programming: How COVID-19 affects software developers and how their organizations can help},
  author={Ralph, Paul and Baltes, Sebastian and Adisaputri, Gianisa and Torkar, Richard and Kovalenko, Vladimir and Kalinowski, Marcos and Novielli, Nicole and Yoo, Shin and Devroey, Xavier and Tan, Xin and others},
  journal={Empirical software engineering},
  volume={25},
  number={6},
  pages={4927--4961},
  year={2020},
  publisher={Springer}
}

@article{godliauskas2025well,
  title={The well-being of software engineers: a systematic literature review and a theory},
  author={Godliauskas, Povilas and {\v{S}}mite, Darja},
  journal={Empirical Software Engineering},
  volume={30},
  number={1},
  pages={35},
  year={2025},
  publisher={Springer}
}

@inproceedings{alzahrani2022impact,
  title={Impact of animated objects on autistic and non-autistic users},
  author={Alzahrani, Mona and Uitdenbogerd, Alexandra L and Spichkova, Maria},
  booktitle={Proceedings of the 2022 ACM/IEEE 44th International Conference on Software Engineering: Software engineering in society},
  pages={102--112},
  year={2022}
}

@article{caldwell2008web,
  title={{Web content accessibility guidelines (WCAG) 2.0}},
  author={Caldwell, Ben and Cooper, Michael and Reid, Loretta Guarino and Vanderheiden, Gregg and Chisholm, Wendy and Slatin, John and White, Jason},
  journal={{WWW Consortium (W3C)}},
  volume={290},
  number={1-34},
  pages={5--12},
  year={2008}
}

@article{borg2011right,
  title={The right to assistive technology: For whom, for what, and by whom?},
  author={Borg, Johan and Larsson, Stig and {\"O}stergren, Per-Olof},
  journal={Disability \& Society},
  volume={26},
  number={2},
  pages={151--167},
  year={2011},
  publisher={Taylor \& Francis}
}

@book{cook2019assistive,
  title={Assistive technologies: Principles \& practice},
  author={Cook, Albert M and Polgar, Janice M and Encarna{\c{c}}{\~a}o, Pedro},
  year={2019},
  publisher={Elsevier}
}

@article{alzahrani2021human,
  title={Human-computer interaction: Influences on autistic users},
  author={Alzahrani, Mona and Uitdenbogerd, Alexandra L and Spichkova, Maria},
  journal={Procedia Computer Science},
  volume={192},
  pages={4691--4700},
  year={2021},
  publisher={Elsevier}
}

@incollection{holz2006research,
  title={Research Methods in Computing: What are they, and how should we teach them?},
  author={Holz, Hilary J and Applin, Anne and Haberman, Bruria and Joyce, Donald and Purchase, Helen and Reed, Catherine},
  booktitle={Working group reports on ITiCSE on Innovation and technology in computer science education},
  pages={96--114},
  year={2006}
}

@inproceedings{weinman2015teaching,
  title={Teaching computing as science in a research experience},
  author={Weinman, Jerod and Jensen, David and Lopatto, David},
  booktitle={Proceedings of the 46th ACM Technical Symposium on Computer Science Education},
  pages={24--29},
  year={2015}
}

@inproceedings{fernandez2024generalizing,
  title={Generalizing the CS Course-based Undergraduate Research Experience (CURE)},
  author={Fernandez, Amanda S},
  booktitle={Proceedings of the 55th ACM Technical Symposium on Computer Science Education V. 2},
  pages={1642--1643},
  year={2024}
}

@inproceedings{host2002introducing,
  title={Introducing empirical software engineering methods in education},
  author={Host, Martin},
  booktitle={Proceedings 15th Conference on Software Engineering Education and Training (CSEE\&T 2002)},
  pages={170--179},
  year={2002},
  organization={IEEE}
}

@article{clunne2017modelling,
  title={Modelling and implementation of humanoid robot behaviour},
  author={Clunne-Kiely, Leroy and Idicula, Bijin and Payne, Luke and Ronggowarsito, Enrico and Spichkova, Maria and Simic, Milan and Schmidt, Heinrich},
  journal={Procedia computer science},
  volume={112},
  pages={2249--2258},
  year={2017},
  publisher={Elsevier}
}

@article{christianto2017software,
  title={Software engineering solutions to support vertical transportation},
  author={Christianto, Alber J and Chen, Peng and Walawedura, Osheen and Vuong, Annie and Feng, Jun and Wang, Dong and Spichkova, Maria},
  journal={arXiv preprint arXiv:1712.04652},
  year={2017}
}

@inproceedings{spichkova2020gosecure,
  title={{GoSecure: Securing Projects with Go}},
  author={Spichkova, Maria and Vaish, Achal and Highet, David C and Irfan, Isthi and Kesley, Kendrick and Kumar, Priyanga D},
  booktitle={15th International Conference on Evaluation of Novel Approaches to Software Engineering (ENASE)},
  year={2020}
}

@article{george2020usage,
  title={Usage visualisation for the AWS services},
  author={George, Lettisia Catherine and Guo, Yanan and Stepanov, Denis and Peri, Vikas Kumar Reddy and Elvitigala, Roshan Lakmal and Spichkova, Maria},
  journal={Procedia Computer Science},
  volume={176},
  pages={3710--3717},
  year={2020},
  publisher={Elsevier}
}

@article{zhao2025visualisation,
  title={Visualisation for the {CIS} benchmark scanning results},
  author={Zhao, Zhenshuo and Spichkova, Maria and Champavat, Duttkumari and Kulkarni, Juilee N and Singla, Sahil and Zulkefli, Muhammad A and Khandelwal, Pradhuman},
  journal={arXiv preprint arXiv:2512.11316},
  year={2025}
}

@inproceedings{spichkova2017autonomous,
  title={Autonomous systems research embedded in teaching},
  author={Spichkova, Maria and Simic, Milan},
  booktitle={International Conference on Intelligent Interactive Multimedia Systems and Services},
  pages={268--277},
  year={2017},
  organization={Springer}
}

@inproceedings{sun2018software,
  title={Software development for autonomous and social robotics systems},
  author={Sun, Chong and Zhang, Jiongyan and Liu, Cong and King, Barry Chew Bao and Zhang, Yuwei and Galle, Matthew and Spichkova, Maria and Simic, Milan},
  booktitle={International Conference on Intelligent Interactive Multimedia Systems and Services},
  pages={151--160},
  year={2018},
  organization={Springer}
}

@article{simic2016enhancing,
  title={Enhancing learning experience by collaborative industrial projects},
  author={Simic, Milan and Spichkova, Maria and Schmidt, Heinrich and Peake, Ian},
  year={2016},
  publisher={RMIT University}
}

\vfill
\end{document}